\documentclass[letterpaper,english,aps,prb,reprint,amsmath,amssymb,onecolumn]{revtex4-2}
\usepackage[T1]{fontenc}
\usepackage[latin9]{inputenc}
\usepackage{color}
\usepackage{babel}
\usepackage{amsmath}
\usepackage{amssymb}
\usepackage{graphicx}
\usepackage{esint}
\usepackage{microtype}
\usepackage[unicode=true,pdfusetitle,
 bookmarks=true,bookmarksnumbered=true,bookmarksopen=true,bookmarksopenlevel=1,
 breaklinks=false,pdfborder={0 0 0},pdfborderstyle={},backref=false,colorlinks=true]
 {hyperref}

\makeatletter

\pdfpageheight\paperheight
\pdfpagewidth\paperwidth

\providecommand{\tabularnewline}{\\}

\usepackage{braket}
\usepackage{mathtools}
\usepackage{graphicx}
\hypersetup{linkcolor = blue,
            urlcolor  = blue,
            citecolor = blue,
            anchorcolor = blue}

\makeatother

\begin{document}
\title{Quantum tunneling of magnetization in molecular spin}
\author{Le Tuan Anh Ho}
\email{chmhlta@nus.edu.sg}

\affiliation{Department of Chemistry, National University of Singapore, 3 Science Drive 3 Singapore 117543}
\author{Liviu Ungur}
\email{chmlu@nus.edu.sg}

\affiliation{Department of Chemistry, National University of Singapore, 3 Science Drive 3 Singapore 117543}
\author{Liviu F. Chibotaru}
\email{liviu.chibotaru@kuleuven.be }

\affiliation{Theory of Nanomaterials Group, Katholieke Universiteit Leuven, Celestijnenlaan 200F, B-3001 Leuven, Belgium}
\date{\today}
\begin{abstract}
We examine the quantum tunneling of magnetization in molecular spin in weak interaction with a bath subject to Redfield master equation. By designing a microscopic model for a multilevel spin system using only a generic Hamiltonian and applying stationary approximation for excited doublets/singlets, we derive a key equation of motion for the quantum tunneling of magnetization process which is applicable in the whole temperature domain. From this equation, we find that in general three tunneling rates are needed to accurately describe the quantum tunneling process. More importantly, behavior of the quantum tunneling in the intermediate temperature domain where there exists a transition between incoherent and coherent quantum tunneling is also unraveled for the first time. Limiting cases at low and high temperature and/or low magnetic field are also worked out where some popular well-known results are reproduced. Last but not least, a new interpretation of the quantum tunneling of magnetization is proposed where we reveal the similarity between this relaxation process with a driven damped harmonic oscillator. 
\end{abstract}
\maketitle
\global\long\def\hmt{\mathcal{H}}%
\global\long\def\vt#1{\mathbf{#1}}%

\global\long\def\chip{\chi'}%
\global\long\def\chipp{\chi''}%

\global\long\def\ketbra{\ket{m}\bra{m}}%

\section{Introduction}

In reality, any quantum system is always in interaction with the environment. This interaction results in a change in the dynamics of the system and leads to the quantum dissipation of either information and/or energy \citep{caldeira2014introduction,Blum1996,banerjee2018open}. Magnetic system such as a spin is not an exception. Via interaction with other degrees of freedom of environment, the spin tries to reach the equilibrium state with its surroundings. This is macroscopically exhibited as a change in the magnetization of the magnetic material sample. Time evolution of this magnetization relaxation process is of utmost importance to understanding a magnetic material since it provides essential knowledge in improving the material magnetic performance \citep{Gatteschi2006,Bartolome2014a,Bartolome2017,Moreno-Pineda2021}.

In general, time evolution of magnetization of a magnetic subsystem can be determined by solving the quantum Liouville equation for the whole environment plus subsystem. This can be simplified into the equation of the reduced density matrix which describes only the change of the subsystem over time under the effect of either thermal or spin bath \citep{Blum1996,Garanin2011,Prokof'ev2000}. Due to the scale and complexity of the environment, many approximations have been invoked. In particular, for a subsystem weakly interacting with a bath, the most popular one is Born-Markov approximation with the corresponding Redfield equation \citep{Blum1996,Garanin2011}. In the case of molecular spin, this equation not only allows to find the magnetization relaxation rate but also help elucidate the role of each constituent relaxation process such as direct process, Orbach process, Raman process, and especially quantum tunneling of magnetization (QTM) \citep{Abragam1970,Gatteschi2006,Bartolome2017} 

Since the emergence of single-molecule magnets nearly three decades ago \citep{Sessoli1993}, great efforts have been devoted to study the phenomenon of quantum tunneling of magnetization in this type of material particularly and molecular spin in general \citep{Garanin1997,Leuenberger2000,Gatteschi2006,Bartolome2017}. Being a molecular spin in interaction with mainly thermal bath, physics of this type of material can also be investigated using the mentioned formalism. Application of the Redfield equation at high temperature reveals that the quantum tunneling of magnetization behaves in an incoherent manner under the effect of the thermal (phonon) bath. A corresponding incoherent quantum tunneling rate was found and ubiquitously used since then \citep{Garanin1997,Leuenberger2000,Gatteschi2003,Gatteschi2006}. Meanwhile, at very low temperature where only the ground states are populated and no decoherence exists, it is well-known that the population oscillates between two localized states as a Rabi oscillation due to the existence of either an intrinsic or field-induced quantum tunneling splitting gap \citep{rabi1937space,Garanin2011,Gatteschi2006}. However, according to our best knowledge, behavior of the quantum tunneling of magnetization in the whole domain of temperature, especially the intermediate temperature domain where there is a transition between incoherent to coherent quantum tunneling and/or low temperature domain where some decoherence and small decaying of magnetization exist, has not been thoroughly considered. 

Hence, while we refer to one of our companion work \citep{Ho2022b}, here we focus on the description of the quantum tunneling of magnetization process of a multilevel spin system $S\left(J\right)$ in weak interaction with a thermal bath. Based on the approach presented in companion work \citep{Ho2022b}, we work out on the quantum tunneling of magnetization phenomenon on the whole domain of temperature and applied magnetic field. Results for several limiting cases  will be also given. Furthermore, we also introduce a new interpretation of the quantum tunneling process in connection with the harmonic oscillator in this work. 

The article is organized as follows. In Section II, we first introduce the microscopic model and the corresponding equation of motion describing the quantum tunneling of magnetization process. In Section III, the general solutions resulting from the equation describing the QTM are given. Some limiting cases are then presented in section IV. Section V is dedicated to a new interpretation of the quantum tunneling of magnetization in connection with the driven and damped harmonic oscillator. A summary on the findings and discussions of their applications are finally given in the last section.

\section{Microscopic description of quantum tunneling of magnetization}

To investigate the quantum tunneling of magnetization process, we consider a spin system $S$ $\left(J\right)$ with the following generic Hamiltonian in the localized basis \citep{Garanin2011,Ho2017,Ho2022a}:

\begin{multline}
\hmt=\sum_{m^{\mathrm{th}}}\left(\varepsilon_{m}+\frac{W_{m}}{2}\right)\ket{m}\bra{m}+\left(\varepsilon_{m}-\frac{W_{m}}{2}\right)\ket{m'}\bra{m'}+\sum_{m^{\mathrm{th}}}\left(\frac{\Delta_{m}}{2}\ket{m}\bra{m'}+\frac{\Delta_{m}^{*}}{2}\ket{m'}\bra{m}\right)+\sum_{n^{\mathrm{th}}}\varepsilon_{n}\ket{n}\bra{n},
\end{multline}
where $m$ ($n$) indicates quantities corresponding to the doublet $m^{\mathrm{th}}$ (singlet $n^{\mathrm{th}}$), $W_{m}$ is the energy bias induced by the magnetic field, and $\Delta_{m}$ is the tunneling splitting gap of the corresponding $m^{\mathrm{th}}$ doublet. This spin system $S$ is supposed to be in weak interaction with a bath and subject to the Redfield equation \citep{Blum1996,Garanin2011}. Using the semi-secular approximation \citep{Garanin2011,Ho2017} for the Redfield equation and the stationary limit for excited doublets/singlets \citep{Ho2022b}, we obtain the equation for the density matrix element of the ground doublet \citep{Ho2022b}:

\begin{gather}
\frac{dX_{1}}{dt}=-\Gamma_{e}X_{1}-2\left(\Delta_{1r}\rho_{11'i}-\Delta_{1i}\rho_{11'r}\right),\label{eq:dX1}\\
\frac{d\rho_{11'r}}{dt}=-\gamma_{11'}\rho_{11'r}+W_{1}\rho_{11'i}-\frac{\Delta_{1i}}{2}X_{1},\label{eq:drho11'r}\\
\frac{d\rho_{11'i}}{dt}=-W_{1}\rho_{11'r}-\gamma_{11'}\rho_{11'i}+\frac{\Delta_{1r}}{2}X_{1},\label{eq:drho11'i}
\end{gather}
where $X_{1}=\rho_{11}-\rho_{1'1'}$ is the population difference between two localized states corresponding to the ground doublet; $\rho_{11'r}$ and $\rho_{11'i}$ are respectively the real and imaginary component of $\rho_{11'}$; and $\Delta_{1r}$ and $\Delta_{1i}$ are the real and imaginary component of the ground doublet tunneling splitting $\Delta_{1}$. Here it is clear that $\Gamma_{e}$ plays the role of the relaxation rate of the ground doublet population difference when there is no tunneling splitting gap in the ground doublet; $\gamma_{11'}$ is the decoherence rate (escape rate) of the ground doublet population \citep{Ho2022b}. It should also be noted that under the stationary limit for excited doublets/singlets , density matrix elements of the excited doublets/singlets are linear combinations of the density matrix elements of the ground doublet and thus subject to the same relaxation behavior of the ground doublet density matrix elements \citep{Ho2022b}. 

By changing the variables $\rho_{11'r}$ and $\rho_{11'i}$ into $\rho_{r}\equiv\left(\Delta_{1i}\rho_{11'i}+\Delta_{1r}\rho_{11'r}\right)/\Delta_{1}$ and $\rho_{i}\equiv\left(\Delta_{1i}\rho_{11'r}-\Delta_{1r}\rho_{11'i}\right)/\Delta_{1}$, Eqs. (\ref{eq:dX1}-\ref{eq:drho11'i}) is transformed into: 
\begin{align}
\frac{dX_{1}}{dt} & =-\Gamma_{e}X_{1}+2\Delta_{1}\rho_{i},\label{eq:dX1_new_var}\\
\frac{d\rho_{i}}{dt} & =-\gamma_{11'}\rho_{i}+W_{1}\rho_{r}-\frac{\Delta_{1}}{2}X_{1},\label{eq:drho_i}\\
\frac{d\rho_{r}}{dt} & =-\gamma_{11'}\rho_{r}-W_{1}\rho_{i},\label{eq:drho_r}
\end{align}
where we have denoted $\Delta_{1}=\sqrt{\Delta_{1r}^{2}+\Delta_{1i}^{2}}$.

Since $\Gamma_{e}$ plays the role of the relaxation rate when there is no tunneling splitting gap in the ground doublet, we separate this from the solution of the above equations by substituting $X_{1}=xe^{-\Gamma_{e}t}$, $\rho_{i}=p_{i}e^{-\Gamma_{e}t}$, $\rho_{r}=p_{r}e^{-\Gamma_{e}t}$ into Eqs. (\ref{eq:dX1_new_var}-\ref{eq:drho_r}) in order to obtain the following system of equations : 
\begin{align}
\frac{dx}{dt} & =2\Delta_{1}p_{i},\label{eq:dx/dt}\\
\frac{dp_{i}}{dt} & =-2\gamma p_{i}+W_{1}p_{r}-\frac{\Delta_{1}}{2}x,\\
\frac{dp_{r}}{dt} & =-2\gamma p_{r}-W_{1}p_{i},\label{eq:dp2/dt}
\end{align}
where $\gamma\equiv\left(\gamma_{11'}-\Gamma_{e}\right)/2$. This results in the key differential equation governing the quantum tunneling of magnetization process:
\begin{gather}
\frac{d^{3}x}{dt^{3}}+4\gamma\frac{d^{2}x}{dt^{2}}+\left(4\gamma^{2}+\Delta_{1}^{2}+W_{1}^{2}\right)\frac{dx}{dt}+2\gamma\Delta_{1}^{2}x=0,\label{eq:tunneling_3rd_order_eq}
\end{gather}

\section{General solutions}

Solution of the above governing equation of the quantum tunneling of magnetization process can be found in the form $x=\sum_{i=1}^{3}c_{i}e^{-\Gamma_{i}t}$ where $\Gamma_{i}$ are solution of the corresponding characteristic equation: 
\begin{equation}
\Gamma^{3}-4\gamma\Gamma^{2}+\left(4\gamma^{2}+\Delta_{1}^{2}+W_{1}^{2}\right)\Gamma-2\gamma\Delta_{1}^{2}=0,\label{eq:characteristic equation}
\end{equation}
which leads to: 
\begin{align}
\Gamma_{1} & =\frac{4\gamma}{3}-\frac{3\Omega_{1}^{2}-4\gamma^{2}}{3S}+\frac{1}{3}S,\label{eq:Gamma_1}\\
\Gamma_{2} & =\frac{4\gamma}{3}+\frac{1+i\sqrt{3}}{6}\frac{3\Omega_{1}^{2}-4\gamma^{2}}{S}-\frac{1-i\sqrt{3}}{6}S,\\
\Gamma_{3} & =\frac{4\gamma}{3}+\frac{1-i\sqrt{3}}{6}\frac{3\Omega_{1}^{2}-4\gamma_{1}^{2}}{S}-\frac{1+i\sqrt{3}}{6}S,\label{eq:Gamma_3}
\end{align}
where 
\begin{align}
\Omega_{1} & =\sqrt{\Delta_{1}^{2}+W_{1}^{2}},\\
S\equiv & \sqrt[3]{9\gamma\left(\Delta_{1}^{2}-2W_{1}^{2}\right)-8\gamma^{3}+3\sqrt{3}\sqrt{D}},\\
D\equiv & 16\gamma^{4}W_{1}^{2}+\gamma^{2}\left(8W_{1}^{4}-20W_{1}^{2}\Delta_{1}^{2}-\Delta_{1}^{4}\right)+\Omega_{1}^{6},
\end{align}

The constants $c_{i}$ certainly depends on the initial conditions. Assuming that at $t=0$, the whole population is at the state $\ket{1}$, we have  $\left(X_{1},\rho_{11'r},\rho_{11'i}\right)|_{t=0}=\left(1,0,0\right)$, it is straightforward to find that $\left(x,dx/dt,d^{2}x/dt^{2}\right)|_{t=0}=\left(1,0,-\Delta_{1}^{2}\right)$ and accordingly, 
\begin{gather}
\sum_{i=1}^{3}c_{i}=1,\quad\sum_{i=1}^{3}\Gamma_{i}c_{i}=0,\quad\sum\Gamma_{i}^{2}c_{i}=-\Delta_{1}^{2},
\end{gather}
which results in: 
\begin{align}
c_{1} & =\frac{\Gamma_{2}\Gamma_{3}-\Delta_{1}^{2}}{\left(\Gamma_{1}-\Gamma_{2}\right)\left(\Gamma_{1}-\Gamma_{3}\right)},\label{eq:c1}\\
c_{2} & =\frac{\Gamma_{1}\Gamma_{3}-\Delta_{1}^{2}}{\left(\Gamma_{2}-\Gamma_{1}\right)\left(\Gamma_{2}-\Gamma_{3}\right)},\\
c_{3} & =\frac{\Gamma_{1}\Gamma_{2}-\Delta_{1}^{2}}{\left(\Gamma_{3}-\Gamma_{1}\right)\left(\Gamma_{3}-\Gamma_{2}\right)},\label{eq:c3}
\end{align}
 where $\Gamma_{i}$, $i=1,2,3$, are given above. Quantum tunneling of the magnetization is then described via the time-dependent population difference of the ground doublet as follows: 
\begin{gather}
M\left(t\right)\propto x\left(t\right)=\frac{\Gamma_{2}\Gamma_{3}-\Delta_{1}^{2}}{\left(\Gamma_{1}-\Gamma_{2}\right)\left(\Gamma_{1}-\Gamma_{3}\right)}e^{-\Gamma_{1}t}+\frac{\Gamma_{1}\Gamma_{3}-\Delta_{1}^{2}}{\left(\Gamma_{2}-\Gamma_{1}\right)\left(\Gamma_{2}-\Gamma_{3}\right)}e^{-\Gamma_{2}t}+\frac{\Gamma_{1}\Gamma_{2}-\Delta_{1}^{2}}{\left(\Gamma_{3}-\Gamma_{1}\right)\left(\Gamma_{3}-\Gamma_{2}\right)}e^{-\Gamma_{3}t}.\label{eq:M(t)-2}
\end{gather}

\section{Limiting cases}

\subsection{At resonance/low magnetic field \label{subsec:At-resonance/low-magnetic}}

We first consider the case when the total longitudinal magnetic field is so small that it only induces a small energy bias $W_{1}\ll\Delta_{1}$. By using the ansatz $\Gamma_{i}=\Gamma_{i}^{0}\left(1+\varepsilon_{i}\right)$ where $\Gamma_{i}^{0}$ are tunneling rates at resonance, we can easily find all above tunneling rates $\Gamma_{i}$. These $\Gamma_{i}^{0}$ and the corresponding $c_{i}^{0}$ can be straightforwardly obtained from solving Eq. \eqref{eq:characteristic equation}: 
\begin{align}
\Gamma_{1}^{0} & =2\gamma,\,c_{1}^{0}=0\\
\Gamma_{2,3}^{0} & =\gamma\pm\sqrt{\gamma^{2}-\Delta_{1}^{2}},\,c_{2,3}^{0}=\frac{1}{2}\mp\frac{\gamma}{2\sqrt{\gamma^{2}-\Delta_{1}^{2}}},\\
M\left(t\right) & \propto x\left(t\right)=\left(\cosh\sqrt{\gamma^{2}-\Delta_{1}^{2}}t+\frac{\gamma}{\sqrt{\gamma^{2}-\Delta_{1}^{2}}}\sinh\sqrt{\gamma^{2}-\Delta_{1}^{2}}t\right)e^{-\gamma t}
\end{align}

Replacing the above $\Gamma_{i}^{0}$ and the ansatz into the characteristic equation \eqref{eq:characteristic equation}, simple approximations to the first order of $W_{1}/\Delta_{1}$ result in $\Gamma_{i}\approx\Gamma_{i}^{0}$, $c_{i}\approx c_{i}^{0}$, and accordingly the same $M\left(t\right)$. In other words, the corrections $\varepsilon_{i}$ due to the effect of the small energy bias $W$ are only of second order of magnitude and can be ignored. 

As can be seen, different from the popular incoherent quantum tunneling rate formula $\Gamma_{\mathrm{incoherent}}^{\mathrm{tn}}=\Delta_{1}^{2}\gamma_{11'}/\left(W_{1}^{2}+\gamma_{11'}^{2}\right)$ which at resonance reduces to $\Delta_{1}^{2}/\gamma_{11'}$ and thus diverges at low temperature where $\gamma_{11'}\ll\Delta_{1}$, our results above instead lead to a Rabi oscillation of the magnetization with frequency $\sqrt{\Delta_{1}^{2}-\gamma^{2}}$ at low temperature, which  describes correctly the behavior of the spin system in this temperature domain.

\begin{figure}
\begin{centering}
\begin{tabular}{cc}
\includegraphics[scale=0.5]{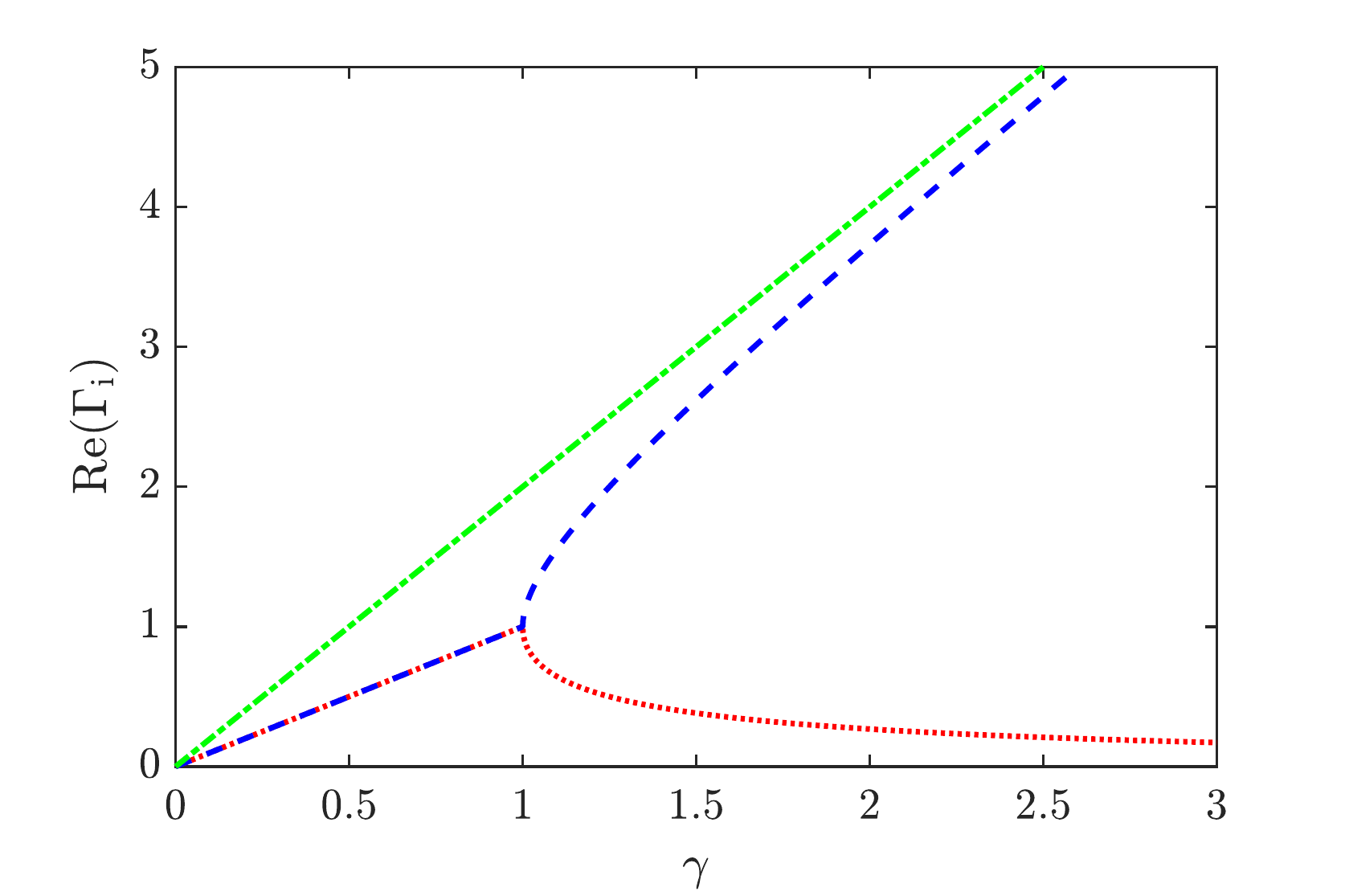} & \includegraphics[scale=0.5]{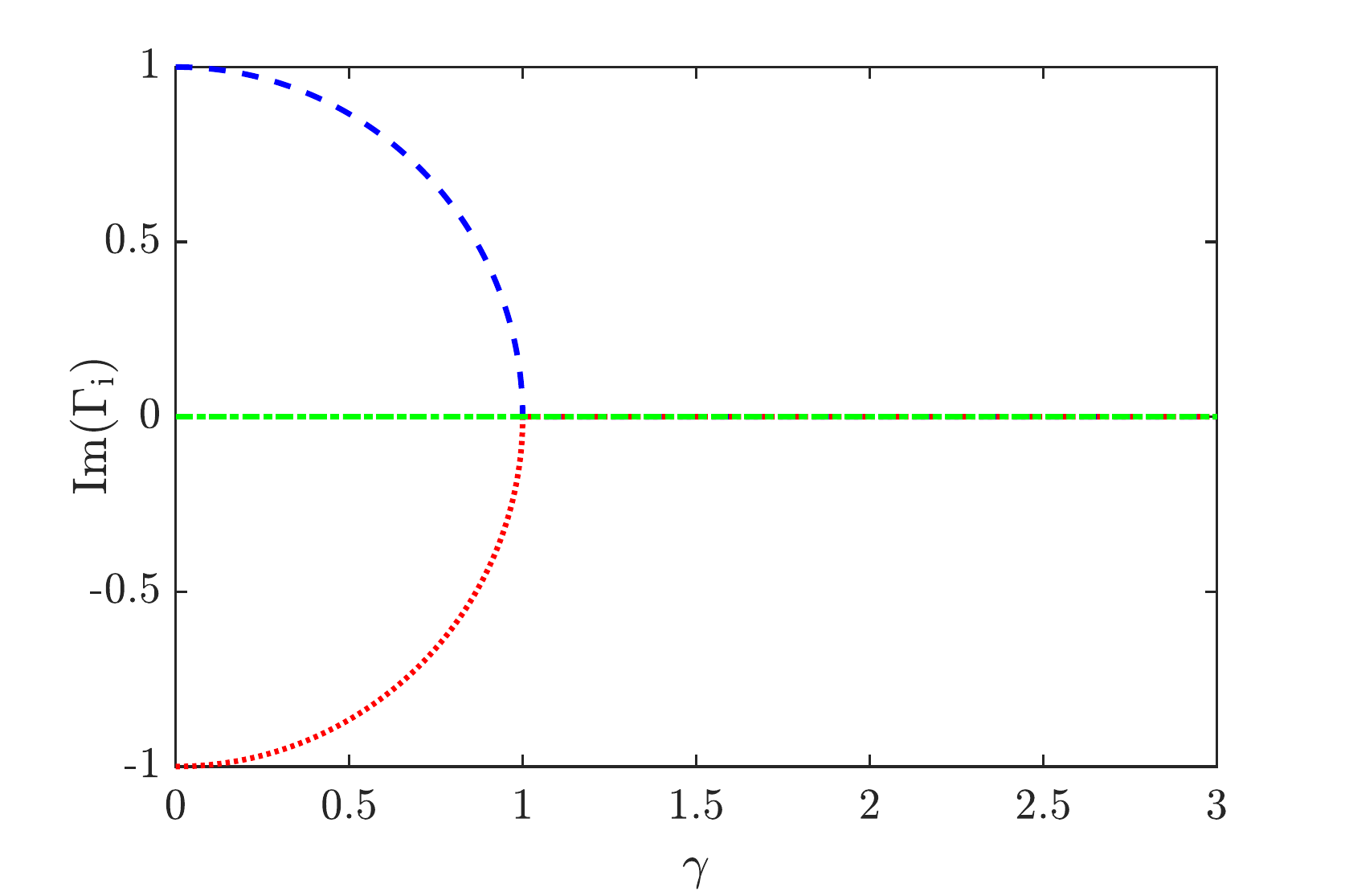}\tabularnewline
\end{tabular}
\par\end{centering}
\caption{Rates of three tunneling mode for $W_{1}=0$ (at resonance) in $\Delta_{1}=1$ unit. \label{fig:Tunneling_rate_W0}}
\end{figure}

It is also worth noticing that the QTM behavior changes when traveling through the special point $\gamma_{0}=\Delta_{1}$. This special point separates two domain where the imaginary part in tunneling rates appear/disappear and can be seen from above equations as well as in Fig. \ref{fig:Tunneling_rate_W0} showing three tunneling rates at resonance. Physically, when $\gamma<\gamma_{0}$ the magnetization oscillates around the equilibrium value with a decaying magnitude. Whereas, the magnetization only decays exponentially for $\gamma>\gamma_{0}$. In other words, the point $\gamma_{0}$ separates the coherent and incoherent QTM. Interestingly, right at $\gamma_{0}$ we have $\Gamma_{1}^{0}=2\Delta_{1}$, $\Gamma_{2,3}=\Delta_{1}$ and the time-dependent magnetization at this special point is thus $M\left(t\right)\propto x\left(t\right)=e^{-\Delta_{1}t}\left(1+\Delta_{1}t\right)$, which aside from the exponential decaying part with time we also have another component increasing linearly with a rate equal to the tunneling splitting $\Delta_{1}$. 

In the case of a large longitudinal magnetic field and accordingly a large energy bias $W_{1}\gg\Delta_{1}$, it is clear that the effect of quantum tunneling is very weak comparing to other relaxation channels. Hence, it is meaningless to discuss about this limiting case.

\subsection{Low temperature \label{subsec:Low-temperature}}

We consider next the limiting case of low temperature domain where the decoherence is much slower than the tunneling frequency between two localized states of the ground doublet, i.e. $\gamma\ll\Delta_{1}$. Simple approximations applied to Eqs. (\ref{eq:Gamma_1}-\ref{eq:Gamma_3}) for three quantum tunneling rates using Taylor series to the first order of $\gamma$ results in: 
\begin{align}
\Gamma_{1} & \approx2\gamma\sin^{2}\theta\\
\Gamma_{2,3} & \approx\left(1+\cos^{2}\theta\right)\gamma\pm i\Omega_{1},
\end{align}
where we have defined: 
\begin{equation}
\sin\theta\equiv\frac{\Delta_{1}}{\Omega_{1}},\,\cos\theta\equiv\frac{W_{1}}{\Omega_{1}}.
\end{equation}
Similarly, we have the followings for the constants $c_{i}$: 
\begin{align}
c_{1} & \approx\cos^{2}\theta,\\
c_{2} & \approx\frac{1}{2}\sin^{2}\theta\left[1+i\frac{\gamma}{\Omega_{1}}\left(1+3\cos^{2}\theta\right)\right],\\
c_{3} & \approx\frac{1}{2}\sin^{2}\theta\left[1-i\frac{\gamma}{\Omega_{1}}\left(1+3\cos^{2}\theta\right)\right],
\end{align}
and accordingly the magnetization relaxation due to quantum tunneling: 
\[
M\left(t\right)\propto x\left(t\right)=\cos^{2}\theta e^{-2\sin^{2}\theta\,\gamma t}+\sin^{2}\theta e^{-\left(1+\cos^{2}\theta\right)\gamma t}\left[\cos\Omega_{1}t+\frac{\gamma}{\Omega_{1}}\left(1+3\cos^{2}\theta\right)\sin\Omega_{1}t\right].
\]

In the limit $\gamma\rightarrow0$, or $T\rightarrow0$, we have $M\left(t\right)\propto x\left(t\right)=\frac{W_{1}^{2}}{\Omega_{1}^{2}}+\frac{\Delta_{1}^{2}}{\Omega_{1}^{2}}\cos\Omega_{1}t$, which is expected for the Rabi oscillation of the magnetization when there is no decoherence.

\subsection{High temperature}

At high temperature, the decoherence occurs much faster than the tunneling frequency, i.e. $\gamma\gg\Delta_{1}$. That is to say, tunneling happens in the incoherent manner. Three tunneling rates $\Gamma_{i}$, Eqs. (\ref{eq:Gamma_1}-\ref{eq:Gamma_3}) can then be straightforwardly approximated as: 

\begin{align}
\Gamma_{1} & =\frac{2\Delta_{1}^{2}\gamma}{W_{1}^{2}+4\gamma^{2}},\label{eq:Gamma_1-1}\\
\Gamma_{2,3} & =2\gamma\left(1-\frac{1}{2}\frac{\Delta_{1}^{2}}{W_{1}^{2}+4\gamma^{2}}\right)\pm iW\left(1+\frac{1}{2}\frac{\Delta_{1}^{2}}{W_{1}^{2}+4\gamma^{2}}\right)\approx2\gamma\pm iW,
\end{align}
 and the corresponding constants $c_{i}$ and $x\left(t\right)$:

\begin{gather}
c_{1}=1-\frac{\left(W_{1}^{2}-4\gamma^{2}\right)\Delta_{1}^{2}}{\left(W_{1}^{2}+4\gamma^{2}\right)^{2}}\approx1,\\
c_{2,3}=\frac{\Delta_{1}^{2}}{2\left(W\mp2i\gamma\right)^{2}}\approx0,\\
M\left(t\right)\propto x\left(t\right)=e^{-\frac{2\Delta_{1}^{2}\gamma}{W_{1}^{2}+4\gamma^{2}}t}
\end{gather}
As can be seen, the contribution of the second and third mode to the relaxation of magnetization are negligible. Tunneling of magnetization can then be described well by the slowest mode, whose rate is similar to the popular incoherent one \citep{Garanin1997,Leuenberger2000,Gatteschi2003,Gatteschi2006} $\Gamma_{\mathrm{incoherent}}^{\mathrm{tn}}=\Delta_{1}^{2}\gamma_{11'}/\left(W_{1}^{2}+\gamma_{11'}^{2}\right)$ except $\gamma_{11'}$ is now replaced by $\gamma_{11'}-\Gamma_{e}$ where the effect of the relaxation via canonical channels has also been taken into the decoherence rate.

\subsection{Coherence/incoherent quantum tunneling transition point \label{subsec:At-transition-point}}

From above, we know that the QTM behaves in an incoherent manner at high temperature where it exponentially decays to the equilibrium without oscillation since only one tunneling mode with the corresponding real rate dominates over other two complex conjugate ones. Meanwhile, at low temperature and low magnetic field, the QTM oscillates with a decaying amplitude due to slow decoherence and there is an involvement of all three tunneling modes where two complex conjugates tunneling modes dominate over the one with the real rate.  This allows us to interpolate on the existence of a critical transition temperature, at which QTM transits between the incoherent and coherent behaviors. Since the temperature implicitly enters the main equation of QTM, Eq. \eqref{eq:characteristic equation}, as a variable within the decoherence rate $\gamma$, what we want to find is a transition decoherence rate $\gamma_{0}$ separating incoherent and coherent QTM, which can be logically defined as the rate $\gamma_{0}$ below which 1) there are two complex conjugate tunneling rates so that an oscillation in the magnetization can exist; and 2) the real part of these two starts becoming smaller than the real rate of the third tunneling mode when temperature decreases, i.e. the oscillation part of $M\left(t\right)$ will decay slower than the one without oscillation and the contribution of the oscillation part dominates. 

\begin{figure}
\begin{centering}
\begin{tabular}{cc}
\includegraphics[scale=0.5]{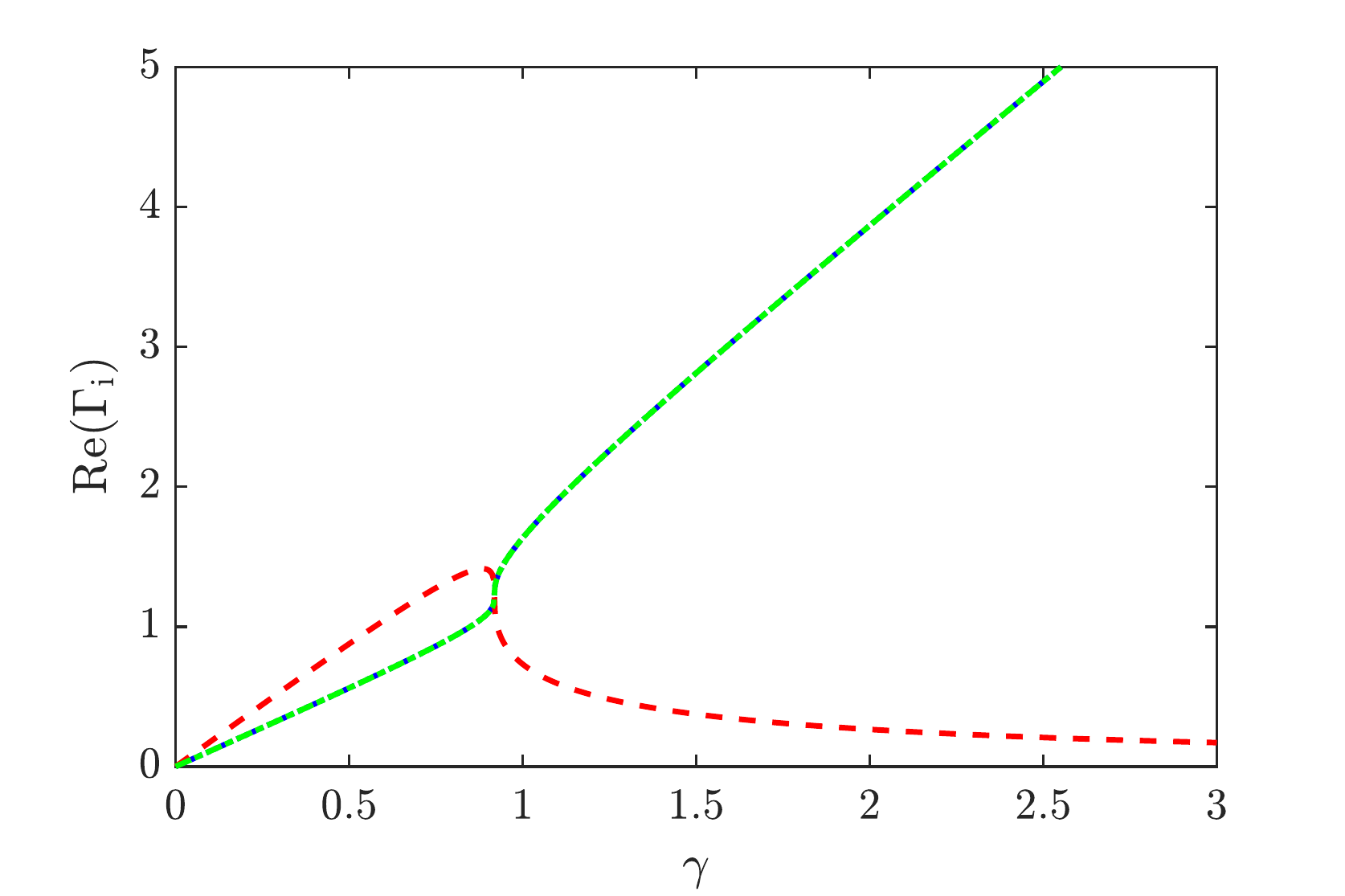} & \includegraphics[scale=0.5]{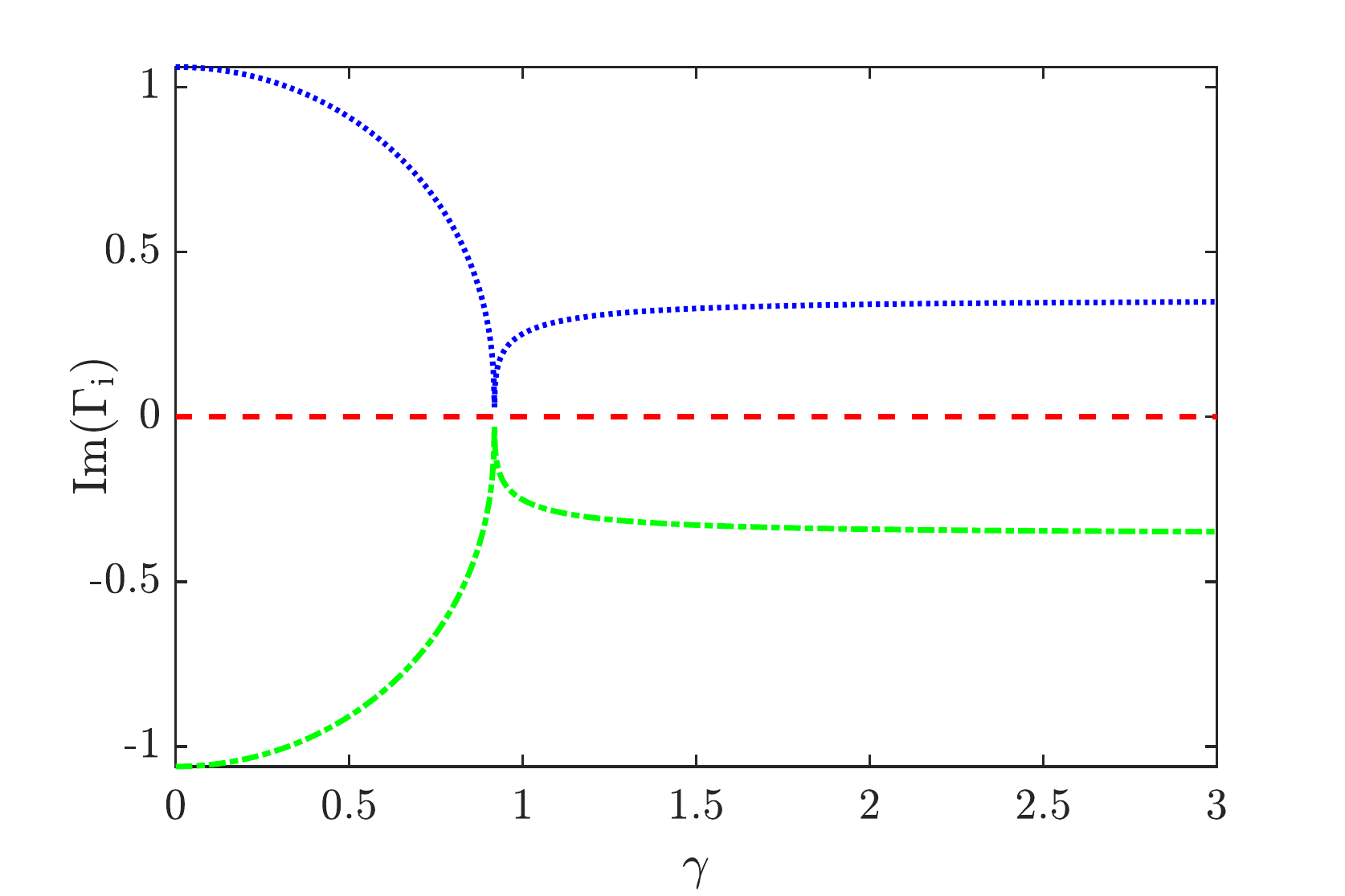}\tabularnewline
\end{tabular}
\par\end{centering}
\caption{Rates of three tunneling mode for $W_{1}=1/2\sqrt{2}$ in $\Delta_{1}=1$ unit. \label{fig:Tunneling_rate_W1}}
\end{figure}

\begin{figure}
\begin{centering}
\begin{tabular}{cc}
\includegraphics[scale=0.5]{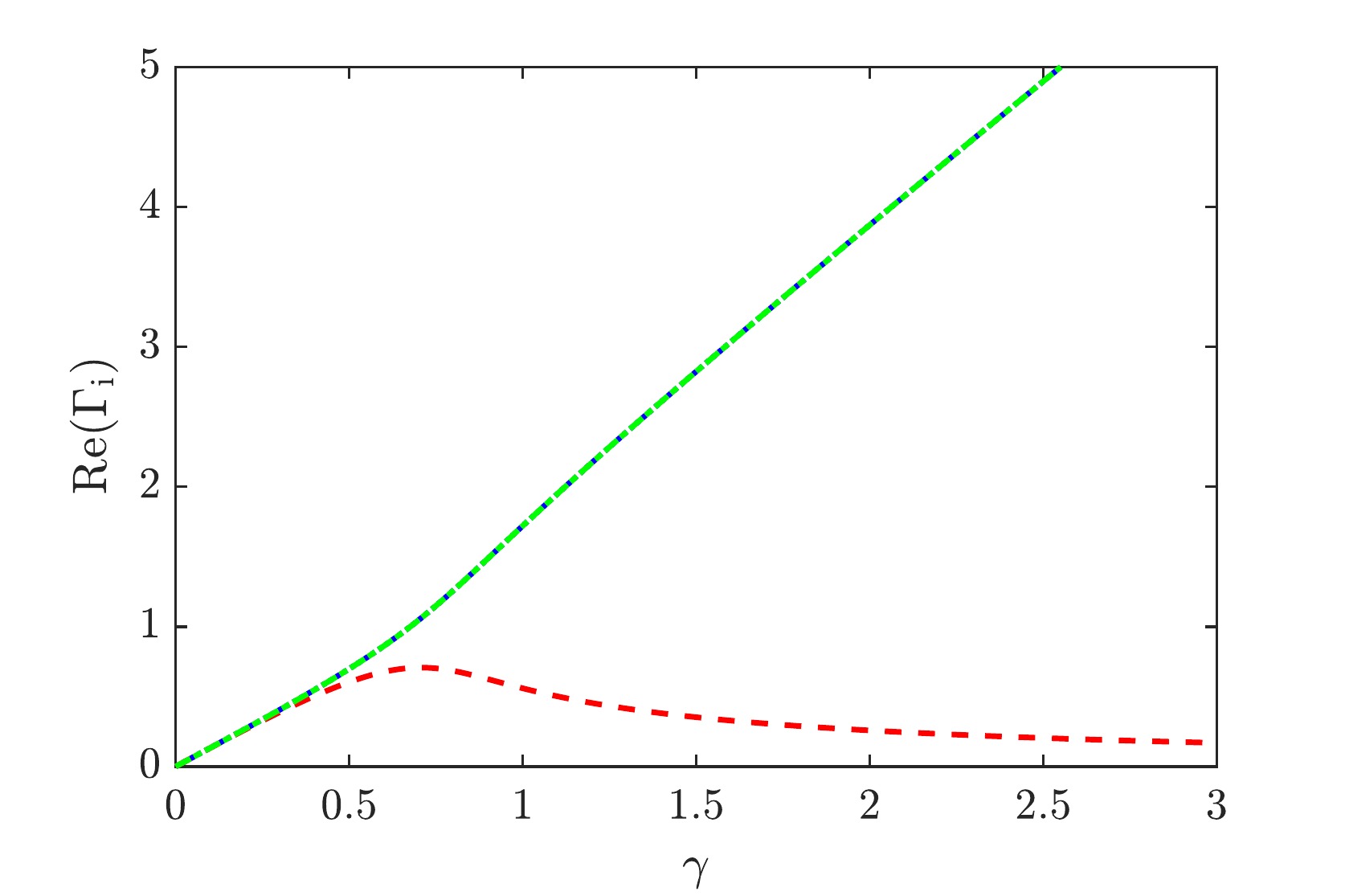} & \includegraphics[scale=0.5]{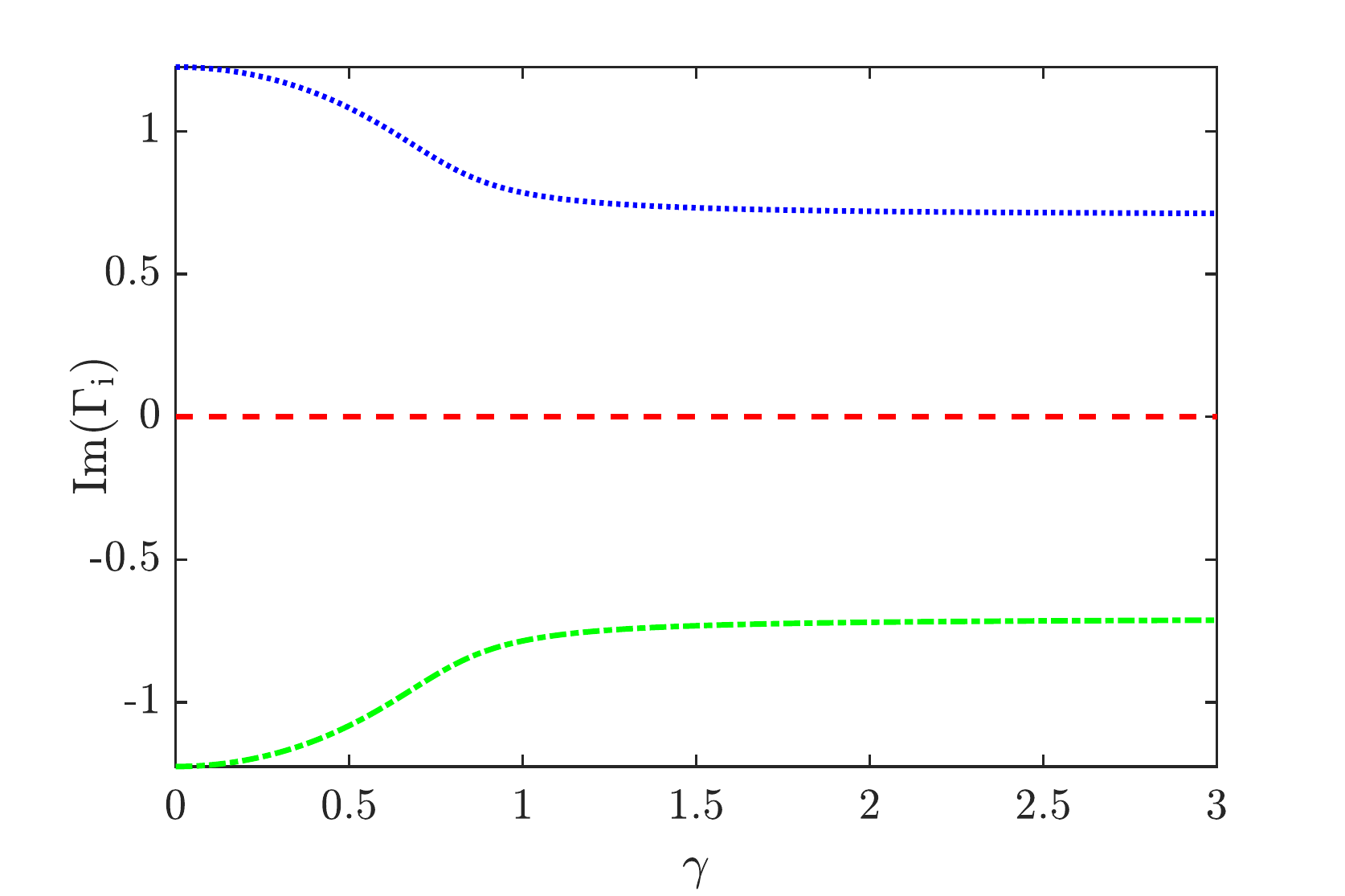}\tabularnewline
\end{tabular}
\par\end{centering}
\caption{Rates of three tunneling mode for $W_{1}=1/\sqrt{2}$ in $\Delta_{1}=1$ unit. \label{fig:Tunneling_rate_W2}}
\end{figure}

Governing equation of QTM, Eq. \eqref{eq:characteristic equation}, is a cubic equation and its solution is thus subject to the Cardano formula. Calculation of the corresponding polynomial discriminant results in a special value of the energy bias $W_{1}=\Delta_{1}/2\sqrt{2}$ beyond which one tunneling rate is always real while the other two are complex conjugate regardless of the value of $\gamma$. This can also be seen from Fig. \ref{fig:Tunneling_rate_W1} and Fig. \ref{fig:Tunneling_rate_W2} which shows three tunneling rates at $W_{1}=\Delta_{1}/2\sqrt{2}$ and $\Delta_{1}/\sqrt{2}$ respectively. In contrast, for $W_{1}<\Delta_{1}/2\sqrt{2}$, the number of real/complex tunneling rates depends not only on $W_{1}$ but also $\gamma$ (see, e.g. Fig. \ref{fig:Tunneling_rate_W0}).  This property thus separates the energy bias $W_{1}$ into two domains where finding the value of the transition decoherence rate $\gamma_{0}$ need different strategies.

\subsubsection{$0\le W_{1}\le\Delta_{1}/2\sqrt{2}$}

\begin{figure}
\begin{centering}
\includegraphics[scale=0.5]{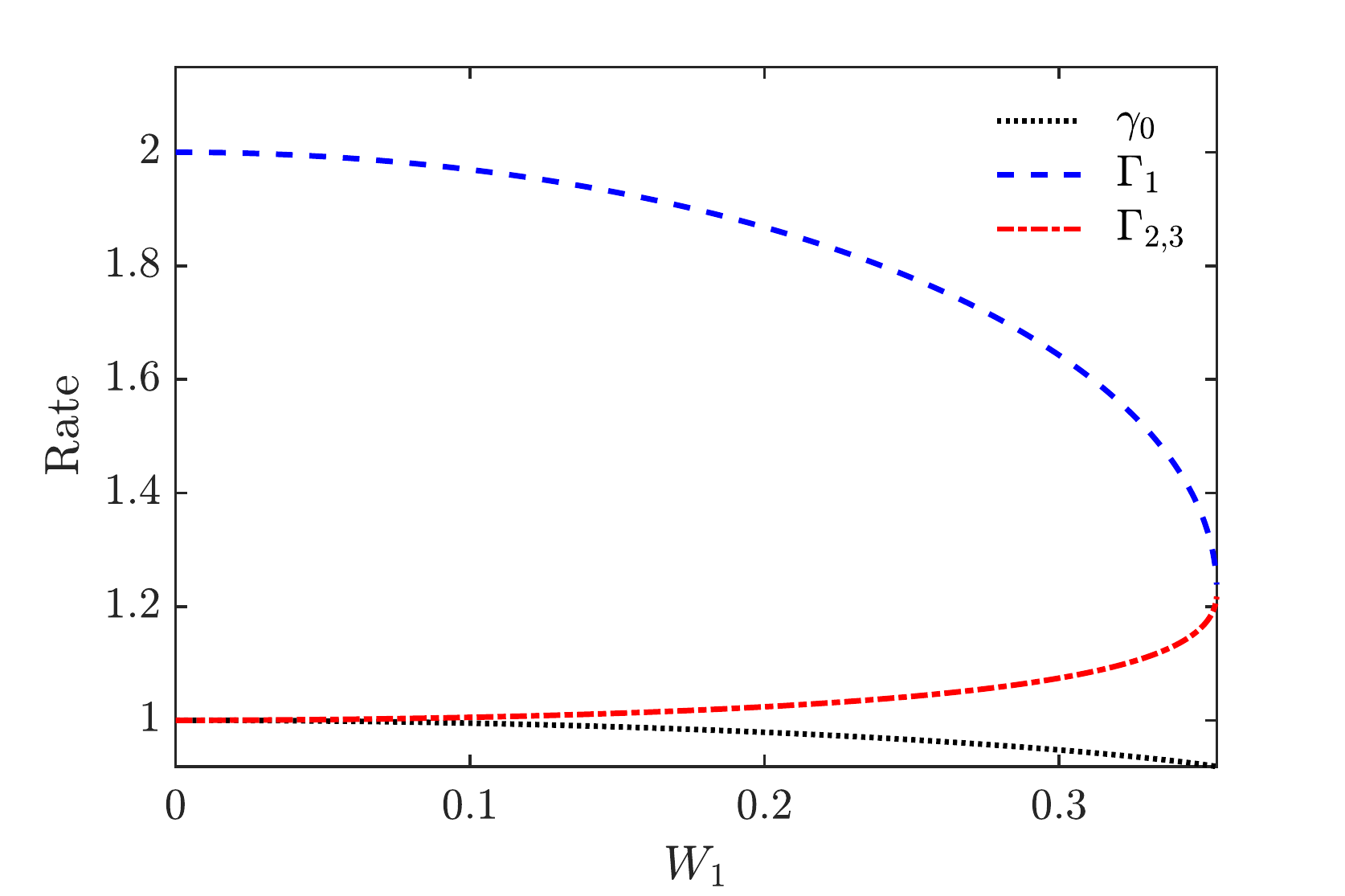}
\par\end{centering}
\caption{Transition point and the corresponding tunneling rates as a function of the energy bias for $0\le W_{1}\le1/2\sqrt{2}$ in $\Delta_{1}=1$ unit. \label{fig:Ttransition point case 1}}
\end{figure}

From previous sections, we know that with a small $\gamma$ and/or low $W_{1}$, there is always two slow complex conjugate tunneling rates and one real faster rate. This two slow tunneling rates will produce the oscillation in QTM. Hence, it is logical that the transition between coherence and incoherence will occurs in the limit when these two slow tunneling rates have their oscillation frequency vanished. Technically, these rates become real-valued and equal, $\Gamma_{1}=\Gamma_{2}$ and $\left\{ \Gamma_{1},\Gamma_{2},\Gamma_{3}\right\} \in\Re$, at the transition point. Using this property, it is straightforward to infer and solve the system of identities corresponding to the coefficients of the Eq. \eqref{eq:characteristic equation}, which then yields the formula for the transition point $\gamma_{0}$: 
\begin{align}
\gamma_{0} & =\frac{\Delta_{1}}{\sqrt{2}}\frac{3+\eta}{4}\sqrt{\frac{3+\eta}{1+\eta}},\label{eq:gamma_0_W small}\\
\eta & \equiv\sqrt{1-8W_{1}^{2}/\Delta_{1}^{2}}
\end{align}
As can be seen, the condition $W_{1}\le\Delta_{1}/2\sqrt{2}$ and accordingly $0\le\eta\le1$ ensures that the transition decoherence rate $\gamma_{0}$ is real and non-negative. The corresponding tunneling rates at this transition point then are: 
\begin{gather}
\Gamma_{1}\left(\gamma_{0}\right)=\sqrt{(1+\eta)(3+\eta)}\frac{\Delta_{1}}{\sqrt{2}},\\
\Gamma_{2}\left(\gamma_{0}\right)=\Gamma_{3}\left(\gamma_{0}\right)=\sqrt{\frac{3+\eta}{1+\eta}}\frac{\Delta_{1}}{\sqrt{2}}.
\end{gather}
These results are depicted in Fig. \ref{fig:Ttransition point case 1} as a function of the energy bias $W_{1}$. As can be seen, within the domain $0\le W_{1}\le\Delta_{1}/2\sqrt{2}$, only decaying rate of faster tunneling modes show a substantial change while the slower mode and the transition point vary little. Additionally, three tunneling rates approach an equal value at the boundary $W_{1}=\Delta_{1}/2\sqrt{2}$.

Due to $\Gamma_{2}\left(\gamma_{0}\right)=\Gamma_{3}\left(\gamma_{0}\right)$ at the transition point, we cannot directly use Eqs. (\ref{eq:c1}-\ref{eq:c3}) to find $c_{i}$ since the solution of the differential equation of QTM, Eq. \eqref{eq:tunneling_3rd_order_eq}, now should be of the form $x\left(t\right)=c_{1}e^{-\Gamma_{1}t}+\left(c_{2}+c_{3}t\right)e^{-\Gamma_{2}t}$ instead of $x\left(t\right)=\sum_{i=1}^{3}c_{i}e^{-\Gamma_{i}t}$. Using the mentioned initial condition $\left(x,dx/dt,d^{2}x/dt^{2}\right)|_{t=0}=\left(1,0,-\Delta_{1}^{2}\right)$, we obtain: 
\begin{gather}
c_{1}=\frac{1-\eta}{\eta}\frac{1}{\eta\left(3+\eta\right)},\,c_{2}=\frac{1+\eta}{\eta}\frac{\eta^{2}+2\eta-1}{\eta\left(3+\eta\right)},\,c_{3}=\frac{1+\eta}{\eta}\sqrt{\frac{1+\eta}{3+\eta}}\frac{\Delta_{1}}{\sqrt{2}},\\
M\left(t\right)\propto x\left(t\right)=\frac{1-\eta}{\eta^{2}\left(3+\eta\right)}e^{-\sqrt{(1+\eta)(3+\eta)}\frac{\Delta_{1}}{\sqrt{2}}t}+\frac{1+\eta}{\eta}\left(\frac{\eta^{2}+2\eta-1}{\eta\left(3+\eta\right)}+\sqrt{\frac{1+\eta}{3+\eta}}\frac{\Delta_{1}}{\sqrt{2}}t\right)e^{-\sqrt{\frac{3+\eta}{1+\eta}}\frac{\Delta_{1}}{\sqrt{2}}t}.
\end{gather}

At resonance $W_{1}=0$, the above results reduce to: 
\begin{gather}
\eta=1,\,\gamma_{0}=\Delta_{1},\\
\Gamma_{1}\left(\gamma_{0}\right)=2\Delta_{1},\,\Gamma_{2}\left(\gamma_{0}\right)=\Gamma_{3}\left(\gamma_{0}\right)=\Delta_{1},\\
c_{1}=0,\,c_{2}=1,\,c_{3}=\Delta_{1},\\
M\left(t\right)\propto x\left(t\right)=\left(1+\Delta_{1}t\right)e^{-\Delta_{1}t},
\end{gather}
which coincides with the result shown in Sec. \ref{subsec:At-resonance/low-magnetic}.

In the limit $W_{1}=\Delta_{1}/2\sqrt{2}$, as can be seen from Fig. \ref{fig:Ttransition point case 1}, all three tunneling rates are equal. Indeed, we obtain: 
\begin{gather}
\eta=0,\,\gamma_{0}=\frac{3}{4}\sqrt{\frac{3}{2}}\Delta_{1},\\
\Gamma_{1}\left(\gamma_{0}\right)=\Gamma_{2}\left(\gamma_{0}\right)=\Gamma_{3}\left(\gamma_{0}\right)=\sqrt{\frac{3}{2}}\Delta_{1},\\
M\left(t\right)\propto x\left(t\right)=\left(1+\sqrt{\frac{3}{2}}\Delta_{1}t+\frac{1}{4}\Delta_{1}^{2}t^{2}\right)e^{-\sqrt{3/2}\Delta_{1}t}.
\end{gather}
Here we have used the solution form $x\left(t\right)=\left(c_{1}+c_{2}t+c_{3}t\right)e^{-\Gamma_{1}t}$ to derive $M\left(t\right)$.

\subsubsection{$W_{1}>\Delta_{1}/2\sqrt{2}$}

In this case, regardless of the value of $\gamma$, one of the tunneling rate (see Eq. \eqref{eq:Gamma_1}) is always real while the others two are complex conjugates. Hence, the first criteria for $\gamma_{0}$ is already satisfied. Noticing that the slowest one is real-valued at high temperature, by decreasing temperature the coherent QTM can be considered starting when the real part of two complex conjugate rates equal to the real rate, i.e. $\Gamma_{2,3}=\Gamma_{1}\pm i\mathrm{Im}\left(\Gamma_{2,3}\right)$ where $\Gamma_{1}\in\Re$. Applying the same method as previously ((finding identities of the coefficients of the characteristic equation), the transition decoherence rate $\gamma_{0}$ can thus be easily found for this case:
\begin{align}
\gamma_{0} & =\frac{3}{2\sqrt{2}}\mu\Delta_{1},\label{eq:gamma_0_W large}\\
\mu & \equiv\sqrt{1-2W_{1}^{2}/\Delta_{1}^{2}}
\end{align}
As can be seen, $\gamma_{0}$ is only meaningful as long as $W_{1}\le\Delta_{1}/\sqrt{2}$. For $W_{1}\ge\Delta_{1}/\sqrt{2}$, the transition decoherence rate $\gamma_{0}$ is right at zero (or temperature goes to 0). 

\begin{figure}
\begin{centering}
\includegraphics[scale=0.5]{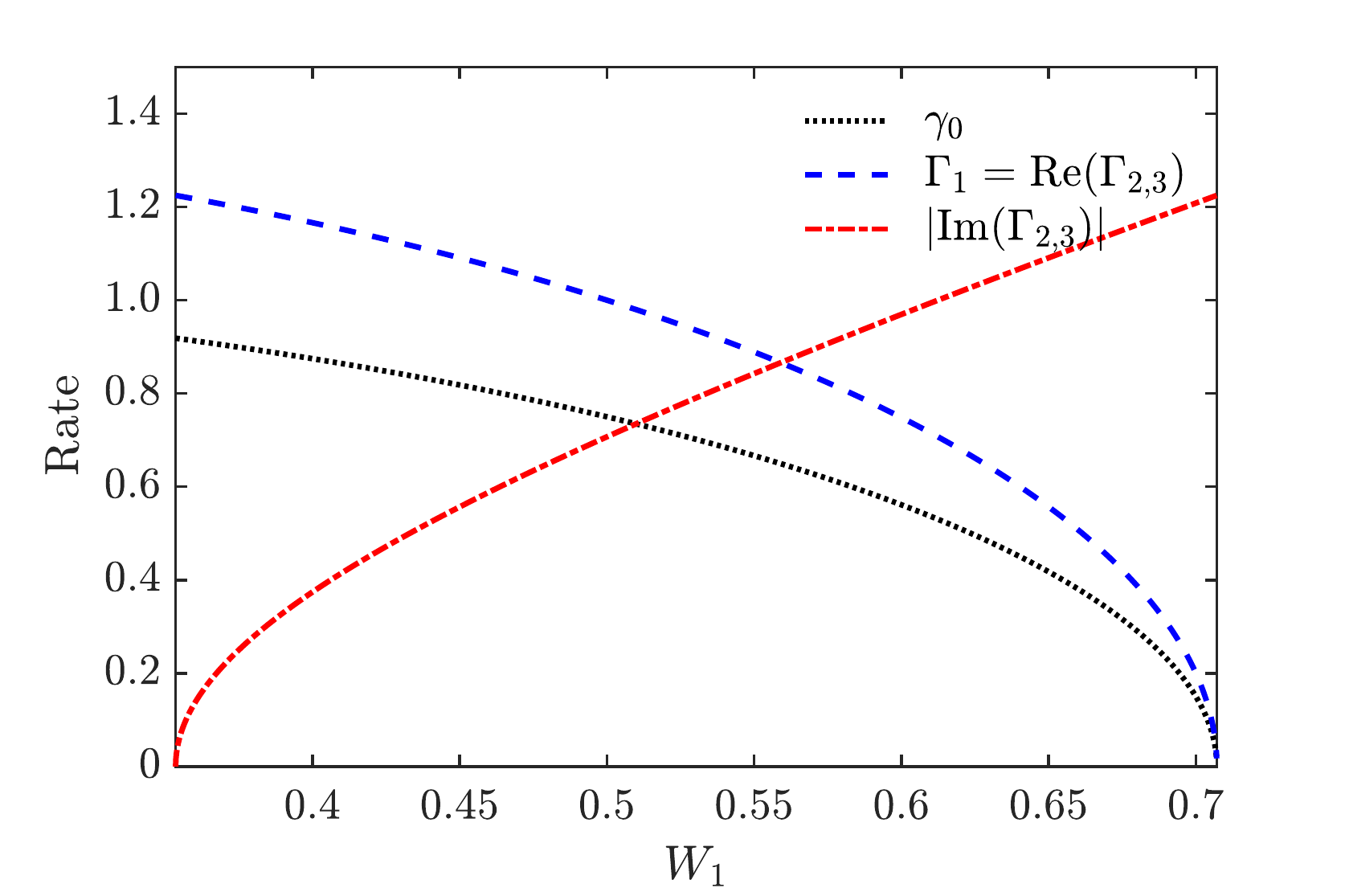}
\par\end{centering}
\caption{Transition point and the corresponding tunneling rates as a function of the energy bias for $W_{1}>1/2\sqrt{2}$ in $\Delta_{1}=1$ unit. \label{fig:Tunneling_rate-2-1-4-1}}
\end{figure}

Given the value of $\gamma_{0}$ and defining: 
\begin{equation}
\nu\equiv\sqrt{8W_{1}^{2}/\Delta_{1}^{2}-1},
\end{equation}
the corresponding $\Gamma_{i}$, $c_{i}$, and $x\left(t\right)$ at $\gamma_{0}$ are: 
\begin{gather}
\Gamma_{1}\left(\gamma_{0}\right)=\sqrt{2}\mu\Delta_{1},\,\Gamma_{2,3}\left(\gamma_{0}\right)=\sqrt{2}\mu\Delta_{1}\pm\frac{i\nu}{\sqrt{2}}\Delta_{1}.\\
c_{1}\left(\gamma_{0}\right)=\frac{1}{\nu^{2}},\,c_{2,3}\left(\gamma_{0}\right)=\frac{1}{2}\left(1-\frac{1}{\nu^{2}}\right)\pm i\frac{\mu}{\nu},\\
M\left(t\right)\propto x\left(t\right)=e^{-\sqrt{2}\mu\Delta_{1}t}\left\{ \frac{1}{\nu^{2}}+\left(1-\frac{1}{\nu^{2}}\right)\cos\left[\frac{\nu\Delta_{1}}{\sqrt{2}}t\right]+2\frac{\mu}{\nu}\sin\left[\frac{\nu\Delta_{1}}{\sqrt{2}}t\right]\right\} .
\end{gather}
These tunneling rates at transition point are also illustrated in Fig. \ref{fig:Tunneling_rate-2-1-4-1} as a function of the energy bias $W_{1}$. The figure clearly shows that in this energy bias domain, both $\gamma_{0}$ and the real part of all $\Gamma_{i}$ decrease rapidly with $W_{1}$ and become 0 at $W_{1}=\Delta_{1}/\sqrt{2}$.

In the limit $W_{1}=\Delta_{1}/\sqrt{2}$, we have: 
\begin{gather}
\mu=0,\,\nu=\sqrt{3},\,\gamma_{0}=0,\\
\Gamma_{1}\left(\gamma_{0}\right)=0,\,\Gamma_{2,3}\left(\gamma_{0}\right)=\pm\sqrt{\frac{3}{2}}i\Delta_{1},\nonumber \\
c_{1}\left(\gamma_{0}\right)=c_{2,3}\left(\gamma_{0}\right)=1/3,\\
M\left(t\right)\propto x\left(t\right)=\frac{1}{3}+\frac{2}{3}\cos\left[\sqrt{\frac{3}{2}}\Delta_{1}t\right].
\end{gather}
Interestingly, three tunneling modes contributes equally to the oscillating QTM in this limit despite the difference in value of the tunneling rates. 

Meanwhile, as $W_{1}=\Delta_{1}/2\sqrt{2}$, we obtain the same results as in the previous section: 
\begin{gather}
\mu=\sqrt{3}/2,\,\nu=0,\,\gamma_{0}=\frac{3}{4}\sqrt{\frac{3}{2}}\Delta_{1},\\
\Gamma_{1}\left(\gamma_{0}\right)=\Gamma_{2}\left(\gamma_{0}\right)=\Gamma_{3}\left(\gamma_{0}\right)=\sqrt{\frac{3}{2}}\Delta_{1},\nonumber \\
M\left(t\right)\propto x\left(t\right)=\left(1+\sqrt{\frac{3}{2}}\Delta_{1}t+\frac{1}{4}\Delta_{1}^{2}t^{2}\right)e^{-\sqrt{3/2}\Delta_{1}t}.
\end{gather}

\section{Quantum tunneling as a driven damped harmonic oscillator}

While Eq. \eqref{eq:tunneling_3rd_order_eq} allow us to find the quantum tunneling rates, it is quite limited in explaining the physics of the quantum tunneling process in the ground doublet. Hence, we will transform Eqs. (\ref{eq:dx/dt}-\ref{eq:dp2/dt}) to serve this purpose. By taking an integration of both sides of Eq. \eqref{eq:dp2/dt} from 0 to $t$ and taking into account the initial condition $\left(x,p_{r},p_{i}\right)|_{t=0}=\left(1,0,0\right)$, we obtain the following equivalent form but intriguing to the understanding of the QTM process: 
\begin{gather}
\frac{d^{2}x}{dt^{2}}+2\gamma\frac{dx}{dt}+\Delta_{1}^{2}x=-W_{1}^{2}\intop_{0}^{t}\mathrm{d}\tau\,e^{-2\gamma\tau}v\left(t-\tau\right),\label{eq:tunneling_damped_harmonic_oscillator}
\end{gather}
where $v\equiv dx/dt$ is the rate of the change of the ground doublet population difference due to QTM.

As can be seen, quantum tunneling of the magnetization in the ground doublet is fundamentally a driven damped harmonic oscillator. In particular, the tunneling splitting $\Delta_{1}$ is the undamped angular frequency of the oscillator. Meanwhile, the rate $\gamma$, which is half of the decoherence rate of the ground doublet states, behaves as the damping coefficient. Accordingly, the damping ratio of this oscillator is $\gamma/\Delta_{1}$. On the other hand, the driving force is proportional to square of the energy bias $W_{1}^{2}$ (or square of the magnetic field $H^{2}$). This driving force is also of the feedback type with some memory effect, which varies according to velocity of the change of the difference in the population of the ground doublet. 

From above, we can also easily see that when the decoherence is very small, the complementary solutions of Eq. 

From the integro-differential equation above, we can also retrieve the familiar incoherent quantum tunneling rate formula. Indeed, when the decoherence rate $\gamma$ is very large, the kernel of the integral in Eq. \eqref{eq:tunneling_damped_harmonic_oscillator} is localized so that the integration $\int_{0}^{t}$ can be expanded into $\int_{0}^{+\infty}$. Supposing that the population relaxation is slow enough so that memory effect is weak, i.e. $v\left(t-\tau\right)\approx v\left(t\right)$, the above equation becomes 
\begin{equation}
\frac{d^{2}x}{dt^{2}}+\left(2\gamma+\frac{W_{1}^{2}}{2\gamma}\right)\frac{dx}{dt}+\Delta_{1}^{2}x=0,
\end{equation}
which leads to the incoherent quantum tunneling rate: 
\begin{equation}
\Gamma^{\mathrm{tn}}=\gamma+\frac{W_{1}^{2}}{4\gamma}-\sqrt{\left(\gamma+\frac{W_{1}^{2}}{4\gamma}\right)^{2}-\Delta_{1}^{2}}\approx\frac{2\Delta_{1}^{2}\gamma}{W_{1}^{2}+4\gamma^{2}}.
\end{equation}

\section{Discussions and conclusions}

In this work, the quantum tunneling of magnetization in molecular spin has been investigated. Only with a generic Hamiltonian and using the stationary limit for excited doublets/singlets, we have derived and solved the key governing equation of motion for this interesting quantum relaxation process, which plays an important role in achieving novel magnetic materials at molecular level. In particular, our work provide a complete description of QTM in whole temperature domain. This spans from high temperature domain where it converges back into the well-known incoherent QTM process, to the intermediate temperature domain where the existence of a transition point for the first time is found and analyzed, and finally at low temperature where a small decaying oscillation of the magnetization is demonstrated or at zero where Rabi oscillation of magnetization occurs. 

From the work has been carried out, it is also important to remind that in general we need up to three tunneling rates for an accurate description of the QTM. This raises a question on the accuracy of experimental works where the QTM process at low temperature was fitted using only one tunneling rates. Another finding of the work is the existence and peculiarity of the transition point where the QTM changes its behavior between coherent and incoherent manner. In particular, at this transition point, the QTM does not purely follow the exponential decaying but there exist another factor increasing with time helps slowing down the relaxation process. Moreover, there may exist some non-monotony in the tunneling rates in the proximity of this special point as well as the discontinuity in the rate of the slowest tunneling mode. We reserve discussion on this peculiarity in the companion paper \citep{Ho2022b}. 

Last but not least, this work also demonstrates that the QTM at its core is a driven damped harmonic oscillator with some memory and feedback effect where the decoherence plays as the damping factor and the magnetic field behaves as an external force. 
\begin{acknowledgments}
L. T. A. H. would like to thank Dr. Naoya Iwahara for helpful discussions. L. T. A. H. and L. U. acknowledge the financial support of the research projects R-143-000-A65-133, A-8000709-00-00, and A-8000017-00-00 of the National University of Singapore. Calculations were done on the ASPIRE-1 cluster (www.nscc.sg) under the projects 11001278 and 51000267. Computational resources of the HPC-NUS are gratefully acknowledged.
\end{acknowledgments}

\bibliographystyle{apsrev4-1}
\bibliography{references}

\end{document}